\documentstyle[12pt,fleqn]{article}
\newcommand{\rf}[1]{\raisebox{1ex}{\cite{#1}}}

\newcommand{\fg}[1]{\item {\label{#1}}}

\begin{document}

\vskip 40pt
   
\centerline {\bf Symbolic dynamics II}
\centerline {\bf  The stadium billiard}
\vskip 1.5cm                                                               
  
\centerline { 
Kai T. Hansen
           }
\centerline {\em Niels Bohr Institute
\footnote{
{\ddag}
{\small Permanent address: Phys Dep., University of Oslo, Box 1048, Blindern,
   N-0316 Oslo}
} 
}
\centerline {\em Blegdamsvej 17, DK-2100 Copenhagen \O}  
\centerline {{\em e-mail:}  khansen@nbivax.nbi.dk}

\vskip 25pt
{\centerline{\bf ABSTRACT}}
\noindent{
We construct a well ordered symbolic dynamics plane for the stadium
billiard.  In this symbolic plane the forbidden and the allowed orbits
are separated by a monotone pruning front, and allowed orbits can be
systematically generated by sequences of approximate finite grammars.
   }
 

\vfill\eject


\section{Introduction} 

The stadium billiard was introduced by Bunimovich\rf{bunimovich} as an
example of a system with non-vanishing Kolmogorov entropy.
The stadium billiard has also been used as an example of a chaotic
quantum system\rf{cqc}. 
The wave functions in the quantized system show complicated
patterns\rf{mcdonald_kaufmann} and the stadium billiard is a model for
which scars are 
conjectured to exist\rf{heller}.  The distribution of quantum
eigenvalues is in good agreement with a GOE
distribution\rf{bohigas_giannoni_schmit}.

The stadium is a convex shaped two dimensional area, limited by two 
half-circles
with radius 1 joined by two straight parallel lines of length $2a$,
fig.~\ref{f_stadium_1}.
A point particle moves with constant velocity in this area and is
reflected elastically when it reaches the border of the area.  The
straight lines can be removed and replaced by an infinite chain of
semicircles as shown in fig.~\ref{f_infinite_stadium}.  The two chains
of semicircles are separated by the distance $2a$ and the same orbits
exist in this billiard as in the stadium.

A symbolic dynamics description of the stadium billiard has not yet
been showed to give an efficient way of computing 
thermodynamic measures. 
For other systems a proper understanding of symbolic
dynamics has  been very useful in semiclassical
calculations\rf{cvitanovic} and we expect the same to apply to
the stadium 
billiard.  There have been several attempts to develop a good symbolic
description of 
all classical orbits in the stadium\rf{biham_kvale,meiss} but the symbolic
description turns out to be more complicated than for other systems,
such as
the logistic map\rf{mss}, the H\'enon map\rf{henon,cgp} and the disk
billiards\rf{hansen_1,hansen_2}.

A symbolic description of a dynamical system assigns to each orbit in the
system a sequence of symbols.  The symbols are elements of a finite or
infinite alphabet.  A covering alphabet assigns distinct symbol sequences
to distinct orbits.
Given a covering alphabet, it still may be true that a given symbol sequence
corresponds to no dynamical orbit in the system; such a sequence is
called a pruned symbol sequence\rf{cgp}.    We shall here distinguish
between pruning that depends on the value of a parameter and pruning
that is parameter independent.
The parameter in the stadium is $a$, the half length of the straight
edge. While in the stadium we have both
kinds of pruning, there also exist systems with
only the parameter dependent pruning.  This is the case for the H\'enon
map.
For sufficiently large parameter $a$, the H\'enon map is a complete
Smale horseshoe\rf{smale} and can be 
described by a binary alphabet without any pruning.  Also the 3-disk
billiard with the distance between the disks sufficiently large can be
described by a complete binary alphabet\rf{hansen_2}.

When the parameter $a$ in the stadium billiard goes to infinity we have
the stadium with only parameter independent pruning, and we shall show 
here which
orbits are forbidden in
this limit.  With decreasing parameter $a$, further orbits are pruned. To
describe this pruning we construct a new {\em well ordered}
alphabet and determine
a {\em pruning front}.  This pruning front is a monotone curve in a
symbol plane and the front changes monotonously with the parameter value.


\section{Covering alphabet for stadium billiard}

The symbolic alphabet is $s_t \in \{
1,2,3,4,5,6\} $ where each symbol corresponds to one bounce of the particle
off a wall in the stadium.  The symbols are indicated in
fig.\ref{f_stadium_1} and are similar to the symbols defined in 
ref.\rf{biham_kvale}:
\begin{enumerate}
  \item A bounce in the lower horizontal line.
  \item A bounce in the upper horizontal line.
  \item A bounce in the right semicircle, anticlockwise with respect to
   the center of this semicircle.
  \item A bounce in the left semicircle, anticlockwise with respect to
   the center of this semicircle.
  \item A bounce in the right semicircle, clockwise with respect to
   the center of this semicircle.
  \item A bounce in the left semicircle, clockwise with respect to
   the center of this semicircle.
\end{enumerate}
We us the conjecture in ref.\rf{biham_kvale} that no two distinct orbits can
be described by the same symbol sequence in this alphabet which has been
tested numerically by Biham and Kvale.

An orbit in the stadium corresponds to an infinite symbol sequence.
The particle bouncing at time $t=0$ with the symbol $s_0$, has the
symbolic description $\cdots s_{-2}s_{-1}s_0s_1s_2 \cdots$.  
Let $x$ be the distance measured along the border of the stadium
anticlockwise from the center point on the lower straight line with
$0\leq x \leq 4 a + 2\pi$.  Let $\phi$
be the angle between the normal vector to the wall and the outgoing
velocity measured anticlockwise $-\pi / 2 \leq \phi \leq \pi /2 $.
The position $x$ and the angle $\phi$ of the bounce at time $t=0$ can be
determined if the orbit exists and we know the infinite symbolic description of
the orbit.  The only 
exception is the periodic orbit $\overline{12} = 
\ldots 121212\ldots$ which has $\phi = 0$ and any position on the
straight lines.

A special case are the orbits going through the center of a semicircle. 
These orbits have neither a clockwise nor an anticlockwise bounce,
so there are two symbols that may represents the same orbit.  The symbolic
past of these orbits are identical with the time reversed symbolic
future, and orbits with this symmetry in the symbols are counted twice.


\section{Parameter independent pruning}

The parameter independent pruning rules are
determined by simple geometrical considerations. 
Biham and Kvale have described some of these rules which they call
geometrical pruning rules in ref.\rf{biham_kvale}.
There cannot be two successive bounces off the same straight line, so
the symbol strings 11 and 22 are forbidden.  A clockwise bounce cannot be
followed by an anticlockwise bounce in the same semicircle and 
a clockwise bounce cannot be followed by an anticlockwise bounce.
This forbids the symbol strings 35, 53, 46 and 64.
These are the only forbidden strings of length two, see
table~\ref{t_geo_pruned}.a.

An orbit that bounces first off the
right semicircle and then off a straight line cannot return to the
right semicircle without bouncing in the left semicircle first.  
This gives rise to a second kind of pruning rules whith strings like 315
and 313 forbidden.  All these forbidden 
right--middle--right and left--middle--left symbol
strings are listed in table~\ref{t_geo_pruned}.b.

The third kind of parameter independent pruned orbits can be
described by using the picture of the chain of semicircles in
fig.~\ref{f_infinite_stadium}.  The figure shows an orbit that first
bounces clockwise on the right side and then bounces clockwise off
a semicircle situated two semicircles below the first one and on the left
side of the billiard.  The same orbit is drawn in the stadium in
fig.~\ref{f_stadium_1} and has the symbolic description $S=5126$. In
fig.~\ref{f_infinite_stadium} it is clear that the outgoing line from the
second bounce has to be above the incoming line since the bounce is
clockwise.  In the symbolic description of the stadium, the string
$S=5126$ cannot be followed by any of the symbol strings:
$\_ 23$, $\_ 25$, $\_ 5$, $\_ 3$, $\_ 13$, $\_ 15$, $\_ 125$,
$\_ 123$,  $\_ 1213$, $\_ 1215$, $\_ 12125$,~\ldots.
More complicated examples of orbits pruned this way can also be
constructed.  This third kind of pruning rules are complicated to state
in symbols $s_t$ while in the well ordered symbols introduced below, the
rule is quite simple.

This third kind of pruning rules also prevents a periodic orbit which
bounces normal to the semicircle to have more than two symbolic 
representations. 
These periodic orbits have an infinite number of bounces normal to the
semicircle but one is however not allowed to choose randomly the symbol
for each normal bounce.  When the symbol for one normal bounce is
choosen there is only one single choice for future and past symbols as
all other choices are pruned by the third kind of parameter independent
pruning rules.


\section{Well ordered alphabet for stadium billiard}

The rule that the symbol strings 11, 22, 35, 46, 53 and 64 are 
forbidden, makes it
possible to construct a new alphabet with 5 symbols.  Each new
symbol is constructed from a combination of two bounces.   
As for any symbol $s$ one of the possible next
symbols is forbidden, only 5 different symbols can follow the symbol $s$.

We demand that the new alphabet is well ordered\rf{hansen_1,hansen_2};
i.e., that there exists
a natural ordering of the symbols which preserves the topological ordering of 
orbits in the phase space $(x,\phi)$.  
Table~\ref{t_alphabet} gives new symbols $v_t \in \{ 0,1,2,3,4\}$, constructed
from all the possible combinations of symbol pairs  $s_{t-1}s_{t}$.  
Table~\ref{t_alphabet} is obtained by observing the change of symbols 
$s_t$ when $\phi$ increases.
Let the bounce be on the lower straight line $s_t=1$.  In the limit $\phi
= -\pi /2$ the next bounce is a anticlockwise bounce in the right
semicircle giving $s_{t+1}=3$ and when $\phi$ increases this bounce
becomes normal to the wall and then clockwise giving $s_{t+1}=5$. 
Increasing $\phi$ further gives $s_{t+1}=2$, then $s_{t+1}=4$ and then
in the end
$s_{t+1}=6$ in the limit $\phi = \pi /2$.  This gives the ordered two
symbol combinations $\{ 13,15,12,14,16\} $ which is the first column in
table~\ref{t_alphabet}.  The same method gives the ordered combinations
for the other bounces.  By fixing the angle $\phi$ and changing the
position $x$ we find that the ordering of two symbol combinations is the
same but we do not necessarily find all the 5 combinations for one value
of $\phi$.  This is a consequence of the pruning. 

As shown in ref.\rf{hansen_2}, the ordering is conserved for a bounce off a
focusing (convex) wall while the ordering is reversed for a bounce off a
straight 
wall.  The reverse ordered symbol is $v_t'=4-v_t \in\{0,1,2,3,4\}$.  The
new well-ordered symbol $w_t$ is equal to $v_t$ or $v_t'$, depending on
whether the 
number of preceeding bounces off the straight edges is odd or even.  If
$p_t$ is the number of symbols 1 and 2 in the sequence $s_0s_1s_2\ldots
s_{t-1}$, the well ordered symbols are given by
\begin{equation}
   w_t = \left\{ \begin{array}{ll}
                  v_t   &  \mbox{if $p_t$ odd} \\
                  4-v_t &  \mbox{if $p_t$ even} \\
                \end{array}
         \right.   \label{a_1}
\end{equation}

We now use the symbols $w_t$ in order to define a real number in base 5 
\begin{equation}
 \gamma = 0.w_1w_2w_3\ldots = \sum_{t=1}^{\infty} \frac{w_t}{5^t} 
          \label{gamma}
\end{equation}
The real number $\gamma$, where $0\leq \gamma \leq 1$, is the symbolic coordinate
representation of the 
future of the orbit starting at the point $(x,\phi)$.

We also need the symbolic past of the orbit.  When we follow an orbit
backward in time, the clockwise and anticlockwise bounces exchange
roles. As our convention we choose the time reversed sequence
$\hat{s}_0\hat{s}_1\hat{s}_2\ldots \hat{s}_t\ldots$  which is related to
the forward symbol sequence $\ldots s_{-t}\ldots s_{-2}s_{-1}s_{0}$ as
follows:

\begin{tabular}{l|llllll}
$s_{-t}$    & 1 & 2 & 3 & 4 & 5 & 6 \\
\hline
$\hat{s}_t$ & 1 & 2 & 5 & 6 & 3 & 4 
\end{tabular} 

With this convention we find that
the values $\hat{w}_t$ are constructed from $\hat{s}_t$ as in
table~\ref{t_alphabet} and we get the real number
\begin{equation}
 \delta = 0.\hat{w}_0\hat{w}_{1}\hat{w}_{2}\ldots = 
        \sum_{t=0}^{\infty} \frac{\hat{w}_{t}}{5^{t+1}}
          \label{delta}
\end{equation}
The coordinate $\delta$ is the symbolic  representation 
of the past of the orbit where $0 \leq \delta \leq 1$.  
The square $(\delta , \gamma )$ is the
symbolic plane representation of the possible stadium billiard trajectories.  
The value of $\gamma $ is
increasing with increasing value of $x$ and with increasing value of
$\phi$.  The value of $\delta $ is increasing with increasing value of
$x$, but decreasing with increasing value of $\phi$.

The phase space for the straight edge and the phase space for the
semicircle
are different so we have to work in two different symbol planes for the
two cases.  The symbol plane for the straight line is denoted
$(\gamma_l,\delta_l)$ and the symbol plane for the semicircle is
denoted $(\gamma_c,\delta_c)$.


\section{Parameter independent pruning in the symbol plane}

The parameter independent pruned orbits discussed above
are easily described in the symbol planes.  The
right--middle--right and left--middle--left forbidden orbits in 
table~\ref{t_geo_pruned}.b are represented in the symbol plane 
$(\gamma_l,\delta_l)$ by all the points in the two squares
$1/2<\gamma_l<1$ and $1/2<\delta_l<1$, and 
$0<\gamma_l<1/2$ and $0<\delta_l<1/2$.
In fig.~\ref{f_orbit_a_5}~a) we plot the points $(\gamma_l,\delta_l)$
corresponding 
to all bounces off a straight line of a particle bouncing $10^5$ times
in the billiard with $a=5$.  As expected, no points are inside the two 
forbidden squares. 

The third kind of parameter independent pruned orbits is easily
described in the semicircle symbol plane.  In the symbol plane for $s_0
= 5$ or $s_0 = 6$, these forbidden orbits are represented as the triangle
$\delta_c \geq \gamma_c + 1/5$.  
The rule follows from comparing a string in the future $s_0s_1s_2\ldots$
with a string in the past $\hat{s}_0\hat{s}_1\hat{s}_2\ldots$.  If the
bounce is 5 or 6 then the symbolic future coordinate should be smaller
than the symbolic past cordinate to reflect that the outgoing angle is 
positive while
the incoming angle is negative.  The limit case is when the angle goes
to 0 (a bounce normal to the wall) where the future symbols
$s_0s_1s_2s_3\ldots$ is identical to the 
past symbols $\hat{s}_0s_1s_2\ldots$, with exception for the first symbol
which is time reversed.  From table~\ref{t_alphabet} we find that this
difference in the first symbol gives a difference in symbolic value
$\delta_c - \gamma_c = 1/5$.  Since the symbolic values are well ordered,
the forbidden strings satisfy
\begin{equation}
   \delta_c \geq \gamma_c + 1/5
\end{equation}
Fig.~\ref{f_orbit_a_5}~b) shows the bounces
of the long chaotic orbit in the symbol plane for $s_t=5$, and as
expected, there are no points above the line $\delta_c = \gamma_c + 1/5$.  If 
$s_t=3$ or $s_t=4$, the symbol plane is the same as for $s_t = 5$ or
$s_t=6$ but reflected with respect to the line $\gamma = \delta$; that
is the same as reversing the time.
In this space the forbidden triangle is given by
\begin{equation}
   \delta_c \leq \gamma_c - 1/5
\end{equation}

The parameter independent pruning can now be summerized as follows:  forbidden
symbol sequences of length 2 (table~\ref{t_geo_pruned}.a) are removed
by introducing the new alphabet; forbidden sequences of the second kind
(table~\ref{t_geo_pruned}.b) are squares in the symbol plane
$(\gamma_l,\delta_l)$;  forbidden sequences of the third kind are
triangles in the symbol plane $(\gamma_c,\delta_c)$.  Unfortunately the
description of the second and the third kind of pruning becomes 
complicated if we try
to describe the second kind in $(\gamma_c,\delta_c)$, or the third kind
in $(\gamma_l,\delta_l)$.


\section{Parameter dependent pruning}
                                                             
Some orbits cannot exist in the stadium if the distance between the two
semicircles is sufficiently short.  These orbits are called
dynamically pruned orbits in ref.\rf{biham_kvale}.
As the parameter $a$ decreases, infinite 
families of orbits are pruned.  We explain the mechanism of this pruning 
by an example.

If  $a>1$ the two periodic orbits $\overline{S_1}=\overline{335553}$ and
$\overline{S_2}=\overline{321521}$ exist and are drawn in
fig.~\ref{f_stadium_with_orbits}~a). If $a=1$, both orbits
bounce in the singular point joining the straight edge and the semicircle, and
are not distinguishable in the phase space, 
see fig.~\ref{f_stadium_with_orbits}~b), even though
the two orbits have different stabilities and different symbolic
description. For 
$a<1$, these two periodic orbits do not exist.  As
shown by Biham and Kvale\rf{biham_kvale}, these forbidden orbits can be
traced by using whole circles instead of the semicircles, and extending
the straight edges beyond the tangencies with  the circles. 
Figs.\ref{f_stadium_with_orbits}~c) and \ref{f_stadium_with_orbits}~d) 
show the two (unphysical) periodic orbits when $a<1$.

This example shows how orbits are pruned when a point in the
orbit reaches the singular point at the junction of the semicircle and
the straight edge. 
The pruning front is determined by finding the symbol sequences of all
orbits bouncing off this singular point, with different angles $\phi$.  In
the symbol planes these orbits give monotone curves as shown in
figs.~\ref{f_prunin_front}~a) and \ref{f_prunin_front}~b).  
The monotonicity follows
directly from the definition of the well ordered symbols.  The two
pruning fronts in the two symbol spaces are obtained by choosing either
the symbol for the straight edge or the symbol for the semicircle as
the symbol $s_0$  describing the bounce off the singular point.
The pruning front is not a continuous curve but a Cantor set of points, with
the gaps in the pruning front generated by the images of the forbidden 
regions in the symbol plane.
Any continuous monotone curve going through all points in the pruning front may
be used as a continuous pruning front.

All bounces of an admissible orbit in the stadium give points on one side of
the pruning front. We refer to the region on the other side of the pruning
front as the {\em primary}
pruned region.  Figs.~\ref{f_orbit_a_1}~a) and \ref{f_orbit_a_1}~b)
show bounces from a long chaotic orbit and a comparison with the
pruning fronts in fig.~\ref{f_prunin_front} shows, as expected, that the
orbit never enters the primary pruned region.

The parameter dependent pruned areas in the symbol plane are:  the
primary pruned region, its images by symmetry, and the images of these
regions by iteration forward and backward in time. Iteration in time is
a shift operation but complicated by the fact that one iterates either 
to the same
symbol plane, or to the other symbol plane, depending on the point.  
A shift operations is always more complicated in well ordered symbols
than in the original symbols\rf{troll}.
One of the symbol plane symmetries is the symmetry between $(\gamma ,\delta)$
and  $(1-\gamma ,1-\delta)$ which follows from the symmetry between odd
and even in the algorithm (\ref{a_1}).  In the $(\gamma_l
,\delta_l)$ symbol plane, 
we also have a symmetry between  $(\gamma_l ,\delta_l)$
and  $(\delta_l ,\gamma_l )$ which arises from the time reversal 
symmetry between an orbit and the time reversed orbit.  This symmetry relates
the point  $(\gamma_c ,\delta_c)$ for bounce with symbol 5 or 6, and the 
point  $(\delta_c ,\gamma_c )$ for a bounce with symbol 3 or 4.  
The symbol describing the bounce off a semicircle changes with time reversal 
since a symbol depends on clockwise or anticlockwise motion.

The primary pruned region can be converted into forbidden symbol
sequences in different approximate ways.  We have choosen to find the
completely forbidden sequences of finite length by the
method described in ref.\rf{hansen_2}.
Table~\ref{t_a_pruned} gives all completely forbidden substrings $S$ of
length less or equal to than 5 for parameter $a=1$.
Longer forbidden substrings can be found from fig.~\ref{f_prunin_front}.

Orbits are pruned only when $a$ decreases, the primary pruned region grows 
monotonously with decreasing $a$, and consequently the topological
entropy is decreasing with $a$.

An orbit is never pruned alone.  An
orbit starting from the singular point can always be represented either
with the symbol for a straight line bounce or a semicircle bounce.  A
periodic orbit bifurcates together with a infinite family of orbits with
similar symbolic description.  The example discussed above gives the
family
\[   \ldots A_1B_15C_1D_14A_2B_25C_2D_24 \ldots \]
\noindent with $A_i\in\{ 1,4\}$, $B_i\in\{ 2,5\}$, $C_i\in\{ 1,5\}$ and
$D_i\in\{ 2,4\}$.   
This family of orbits exists only when $a\geq
1$.  At $a=1$ all the points $(\gamma ,\delta)$ for each orbit in this
family are on the pruning front and this gives a constraint on the
pruning front.  Each orbit bifurcates in a family like this at a singular
parameter value.  In ref.\rf{hansen_per_doubl} we show that for
dispersive billiards the singular bifurcation in a billiard can be
related to a bifurcation tree in a smooth potential and we expect this
to be true also for a focusing billiard, such as the stadium.


\section{Conclusion}

We have shown that the symbolic alphabet introduced in
ref.\rf{biham_kvale} enable us to decide if an orbit is admissible or
not.  This is accomplished by constructing new symbols that
define a well ordered symbol plane.  Meiss\rf{meiss} have showed that
those orbits never bouncing off the straight lines are ordered in 
space by a rotation number, while we have showed that all unstable orbits in
the stadium are ordered in space by using the right symbolic dynamics.
In the symbol plane there are a few connected
regions that are forbidden, and an orbit is admissible if and only if no
bounce of the orbit has a point in the forbidden regions.
We expect that for no parameter values there is a 
complete alphabet or a finite Markov partition,
but sequences of approximate Markov partition can be constructed using the
methods presented here.


\vfill\eject


\begin{enumerate}

\fg{f_stadium_1}
The stadium billiard and the 6 symbols $s_t$.
Symbols 3 and 4 correspond to anticlockwise while 5 and 6 correspond 
to clockwise bounces.
The thin lines trace an orbit with symbolic dynamics $\ldots 5126
\ldots$.

\fg{f_infinite_stadium}
The unfolded stadium with two infinite chains of semicircles.
The thin lines are the same orbit as in fig.~1.

\fg{f_orbit_a_5}
All the bounces of a chaotic orbit bouncing $10^5$ times in the stadium
with parameter $a=5$
plotted in the symbol plane. \ \ a) The symbol plane for the straight
edges. \ \ b) The symbol plane for the clockwise semicircular edges.

\fg{f_stadium_with_orbits}
The two orbits $\overline{335553}$ (dotted line) and $\overline{321521}$ 
(solid line) in the stadium.
\ \ a) The two legal orbits, $a=2$.
\ \ b) The two orbits at the pruning point, $a=1$.
\ \ c) The two forbidden orbits for $a=0.6$.
\ \ d) Blow-up of the upper right corner of c) showing the forbidden bounce
off the straight edge.

\fg{f_prunin_front}
The pruning front in the symbol plane for parameter $a=1$.
\ \ a) The symbol plane for the straight edge.
\ \ b) The symbol plane for the clockwise semicircular edge.

\fg{f_orbit_a_1}
All bounces of a chaotic orbit bouncing $30,000$ times in the stadium
with parameter $a=1$,
plotted in the symbol plane. \ \ a) The symbol plane for straight
edges. \ \ b) The symbol plane for the clockwise semicircular edge.

\end{enumerate}

\vfill\eject

{\bf Acknowledgements: }
The author is grateful to Predrag Cvitanovi\'c and members and guests
of the NBI chaos group at for discussions, to S. Tomsovic for 
making his periodic orbits available, and to the referee for good
sugestions.
The author thanks the Norwegian Research Council for support.


\renewcommand{\baselinestretch} {1}


\begin{table}
\mbox{
$
\rule{0mm}{1cm}
\begin{array}{|c|}
\hline
s_{t}s_{t+1} \\
\hline
11 \\
22 \\
35 \\
53 \\
46 \\
64 \\
\hline
\end{array}
\rule{4mm}{0mm}
\begin{array}{|c|}
\hline
s \\
\hline
3 2^i (1 2)^k 1^j 3 \\
3 2^i (1 2)^k 1^j 5 \\
5 2^i (1 2)^k 1^j 3 \\
5 2^i (1 2)^k 1^j 5 \\
4 2^i (1 2)^k 1^j 4 \\
4 2^i (1 2)^k 1^j 6 \\
6 2^i (1 2)^k 1^j 4 \\
6 2^i (1 2)^k 1^j 6 \\
\hline
\end{array}
$
}

\vspace{5mm}
\mbox{
\hspace{5mm}
a)
\hspace{20mm}
b)
}
\caption{
\label{t_geo_pruned} 
Two kind of symbol sequences that are always forbidden in the stadium
billiard, $i,j \in \{ 0,1\}$ and $k\in \{ 0,1,2,\ldots \}$.
}
\vspace{5cm}
\end{table}

\vfill\eject


\begin{table}
\mbox{
$
\begin{array}{|c|l|}
\hline
s_{t-1}s_{t} &  v_{t} \\
\hline
13 & 0 \\
15 & 1 \\
12 & 2 \\
14 & 3 \\
16 & 4 \\
\hline
\end{array}
\rule{3mm}{0mm}
\begin{array}{|c|l|}
\hline
s_{t-1}s_{t} &  v_{t} \\
\hline
24 & 0 \\
26 & 1 \\
21 & 2 \\
23 & 3 \\
25 & 4 \\
\hline
\end{array}
\rule{3mm}{0mm}
\begin{array}{|c|l|}
\hline
s_{t-1}s_{t} &  v_{t} \\
\hline
33 & 0 \\
32 & 1 \\
34 & 2 \\
36 & 3 \\
31 & 4 \\
\hline
\end{array}
\rule{3mm}{0mm}
\begin{array}{|c|l|}
\hline
s_{t-1}s_{t} &  v_{t} \\
\hline
44 & 0 \\
41 & 1 \\
43 & 2 \\
45 & 3 \\
42 & 4 \\
\hline
\end{array}
\rule{3mm}{0mm}
\begin{array}{|c|l|}
\hline
s_{t-1}s_{t} &  v_{t} \\
\hline
52 & 0 \\
54 & 1 \\
56 & 2 \\
51 & 3 \\
55 & 4 \\
\hline
\end{array}
\rule{3mm}{0mm}
\begin{array}{|c|l|}
\hline
s_{t-1}s_{t} &  v_{t} \\
\hline
61 & 0 \\
63 & 1 \\
65 & 2 \\
62 & 3 \\
66 & 4 \\
\hline
\end{array}
$
}
\caption[Well ordered alphabet]{
\label{t_alphabet}  Construction of the well ordered alphabet in the
stadium billiard.  The well
ordered  symbols $w_t$ are constructed by choosing $w_t=v_t$ when
the number of 1's and 2's (bouncing in a straight line) in the preceeding
symbol string (including $s_{t-1}$) is {\em odd} and choosing
$w_t=4-v_t$ when the number is {\em even}. 
}
\end{table}

\vfill\eject


\begin{table}
\mbox{
$
\begin{array}{|c|}
\hline
w_{-1}w_{0}w_{1}w_{2} \\
\hline
44\cdot 22 \\
44\cdot 21 \\
44\cdot 20 \\
34\cdot 22 \\
34\cdot 21 \\
24\cdot 44 \\
14\cdot 44 \\
04\cdot 44 \\
24\cdot 43 \\
14\cdot 43 \\
\hline
\end{array}
\rule{5mm}{0mm}
\begin{array}{|c|}
\hline
s_{-2}s_{-1}s_{0}s_{1}s_2 \\
\hline
44121 \\
44421 \\
66212 \\
66612 \\
55121 \\
55521 \\
33212 \\
33312 \\
\hline
\end{array}
\begin{array}{|c|}
\hline
\\
\hline
44123 \\
44423 \\
66215 \\
66615 \\
55126 \\
55526 \\
33214 \\
33314 \\
\hline
\end{array}
\begin{array}{|c|}
\hline
\\
\hline
44125 \\
44425 \\
66213 \\
66613 \\
55124 \\
55524 \\
33216 \\
33316 \\
\hline
\end{array}
\begin{array}{|c|}
\hline
\\
\hline
24121 \\
24421 \\
16212 \\
16612 \\
25121 \\
25521 \\
13212 \\
13312 \\
\hline
\end{array}
\begin{array}{|c|}
\hline
\\
\hline
24123 \\
24423 \\
16215 \\
16615 \\
25126 \\
25526 \\
13214 \\
13314 \\
\hline
\end{array}
$
}

\vspace{5mm}

\hspace{5mm} a) \hspace{25mm} b)

\caption{
\label{t_a_pruned}
The parameter dependent completly pruned subsequences of length $\leq 5$ for
parameter $a=1$.
a) Symbols $w_t$.
b) Symbols $s_t$.
}
\end{table}


\begin{thebibliography}{99}                                                   
{\small                                                                       



\bibitem{bunimovich} L. Bunimovich, 
	{\em Funkts. Anal. Ego Prilozh. \bf 8}, 73 (1974);
        L. Bunimovich, 
	{\em Comm. Math. Phys.\bf 65}, 295 (1979)


\bibitem{cqc}P. Cvitanovi\'c, I. Percival and A. Wirzba ed., 
Proceedings of ``Quantum Chaos --
Theory and Experiment'' NATO - ARW ,Copenhagen  28 May  --  1 June  1991;
and {\em Chaos}, periodic orbits issue, {\bf 2}, no 1 (1992).

\bibitem{mcdonald_kaufmann}S. W. McDonald and A. N. Kaufmann
	{\em Phys. Rev. Lett.} {\em 42}, 1189 (1979).

\bibitem{heller} E.J. Heller, {\em Phys.Rev.Lett.} {\bf 53}, 1515 (1984).

\bibitem{bohigas_giannoni_schmit}O. Bohigas and M. J. Giannoni,
	{Lect. Not. Phys.} {\bf 209}, 1 (1984).


\bibitem{cvitanovic} R. Artuso, E. Aurell and P. Cvitanovi\'c,
        {\em Nonlinearity \bf 3}, 325 (1990);
        {\em Nonlinearity \bf 3}, 361 (1990)

\bibitem{biham_kvale}O. Biham and M. Kvale, {\em Characterization of
   Chaotic Billiards Through Periodic Orbits}, Preprint.

\bibitem{meiss}J. D. Meiss,
     {\em Regular Orbits for the Stadium Billiard}
     in ref.~2.a.
     and {\em Chaos}, {\bf 2}, 267 (1992).

\bibitem{mss} N. Metropolis,  M.L. Stein and P.R. Stein, {\em J. Comb. Theo. }
		  {\bf A15}, 25 (1973)

\bibitem{henon} M. H\'enon, {\em Comm. Math. Phys. \bf 50 }, 69 (1976)

\bibitem{cgp}P. Cvitanovi\'{c}, G. H. Gunaratne, I. Procaccia,
        Phys. Rev. A {\bf 38}, 1503 (1988).


\bibitem{hansen_1}K. T. Hansen, {\em Chaos J.} {\bf 2}, 71 (1992).

\bibitem{hansen_2}K. T. Hansen, {\em Symbolic dynamics I},
    Submitted to Nonlinearity.

\bibitem{smale} S. Smale, 
    {\em Bull. Am. Math. Soc.} {\bf 73}, 747, (1967).

\bibitem{hansen_per_doubl}K. T. Hansen, {\em Symbolic dynamics III},
    Submitted to Nonlinearity.

\bibitem{troll} G. Troll,
    Pysica D {\bf 50}, 276 (1991)

} 
\end{thebibliography}
\end{document}